\documentclass[aps,amsmath,prb,twocolumn,a4paper,showpacs]{revtex4}
\usepackage{graphicx}
\begin{document}

\newcommand{\figureheight}{5.2 cm}
\newcommand{\putfig}[2]{\begin{figure}[h]
        \special{isoscale #1.bmp, \the\hsize \figureheight}
        \vspace{\figureheight}
        \caption{#2}
        \label{fig:#1}
        \end{figure}}

\newcommand{\figurehaight}{9.2 cm}
\newcommand{\patfig}[2]{\begin{figure}[h]
        \special{isoscale #1.bmp, \the\hsize \figurehaight}
        \vspace{\figurehaight}
        \caption{#2}
        \label{fig:#1}
        \end{figure}}

\newcommand{\eqn}[1]{(\ref{#1})}
\newcommand{\be}{\begin{equation}}
\newcommand{\ee}{\end{equation}}
\newcommand{\bea}{\begin{eqnarray}}
\newcommand{\eea}{\end{eqnarray}}
\newcommand{\bean}{\begin{eqnarray*}}
\newcommand{\eean}{\end{eqnarray*}}
\newcommand{\nn}{\nonumber}





\title{ \bf  Spin-orbit coupling in a Quantum Dot at high magnetic field}
\author{S. Bellucci $^1$ and P. Onorato $^1$ $^2$ \\}
\address{
        $^1$INFN, Laboratori Nazionali di Frascati,
        P.O. Box 13, 00044 Frascati, Italy. \\
        $^2$Dipartimento di Scienze Fisiche,
        Universit\`{a} di Roma Tre, Via della Vasca Navale 84,
00146 Roma, Italy}
\date{\today}

\begin{abstract}
We describe the simultaneous effects of the spin-orbit (SO)
perturbation and a
 magnetic field $B$ on a disk
 shaped quantum dot (QD). {As it is known the}
  combination of electrostatic forces among the  $N$ electrons
confined in the QD and the Pauli principle can induce a spin
polarization when $B$  (applied in the
direction orthogonal to the QD) is above a threshold value.

In the presence of
 an electric field parallel to $B$, coupled to  the spin $ S $ by a Rashba
 term, we demonstrate that a symmetry breaking takes place:  we
 can observe it by analyzing
 the splitting of the
 levels belonging to an unperturbed  multiplet.
 We also discuss the competitive effects of the magnetic field, the SO perturbation  and
 the electron electron interaction, in order to define the hierarchy of the states belonging to  a
 multiplet. We demonstrate how this hierarchy depends on the QD's size.
 We show the  spin texture due to the combined effects of
 the Rashba effect and the
 interaction responsible for the polarization.


\end{abstract}

\pacs{73.21.La,71.15.Mb,75.75.+a}

\maketitle

\section{Introduction}
The physics of mesoscopic devices  has attracted a lot of
interest in the last two  decades. In particular the electronic
transport properties of semiconducting  Quantum Dots (QDs) were
largely investigated.
The QDs, in which we are interested, are two-dimensional  man-made
"droplets" of charge, confined to a small area within a two
dimensional electron gas (2DEG), which  can contain anything from
a single electron
 to a collection of several thousand ones.
Their typical dimensions range from a few nanometers to a few
microns and their size,
 shape and interactions can be precisely
controlled through the use of advanced nanofabrication
technologies \cite{kou}. The electronic transport through these
devices was  intensely  studied, and it was demonstrated that electron
electron interactions play a central role. In particular, the conductance
of a vertical QD was
measured, which allowed one to
carry out a detailed study of the ground state and the first
excited states of a few electrons confined in the QD.  The
QDs that we have in mind in this paper, are the vertical ones
obtained by an electrostatic confinement due to the gate voltages
as the ones in the experimental set of
ref.\cite{klein,oosterkamp}.

\

The development of the mesoscopic physics favored  the idea of
using the electron spin in these devices, both for transmitting
and processing information\cite{spintro,wolf}. Datta and
Das~\cite{Datta} in 1990 described how the electrical field can be
used to modulate the current and showed the essential role which
the field-dependent spin-orbit (SO) coupling plays in this
mechanism. In semiconductors heterostructures, where a 2DEG is
confined in a potential well\cite{nota} along the $z$ direction,
the SO interaction is of the type proposed by Rashba \cite{[7]}
and Dresselhaus \cite{[8]}: it arises from the asymmetry of the
confining potential which occurs in the physical realization of
the 2DEG,  due to the band offset between $AlGaAs$ and $GaAs$.
 Even
though the Rashba spin splitting is expected to  be very small,
nonetheless this perturbation can give rise to a sizeable modification
of a semiconductor band structure \cite{Stormer,Nitta}.

%
%

The SO interaction comes from the expansion quadratic in $v/c$ of
Dirac equation~\cite{Thankappan} and  is due to the Pauli coupling
between the spin momentum of an electron and a magnetic field,
which appears in the rest frame of the electron, due to its motion
in the electric field. {It follows} that the effects of an
electric field (${\bf E}({\bf R})$ where ${\bf R}$ is the 3D
position vector) on a moving electron have to be analyzed starting
from the following hamiltonian\cite{morozb}:
\begin{equation}
\hat H_{SO} = -\frac{\hbar}{(2M_0c)^2}\;{\bf E}({\bf R})
\left[\hat{{\bf \sigma}}\times \left\{\hat{\bf p}-\frac{e}{c}{\bf
A}({\bf R})\right\}\right]. \label{H_SO}
\end{equation}
Here $M_0$ is the free electron mass, 
 $\hat{{\sigma}}$ are
the Pauli matrices, ${\bf A}$ is the vector potential.

The interface electric field (${\bf E}({\bf R})\approx(0,0,E_z)$),
which accompanies the quantum well asymmetry  in  semiconductors
heterostructures, is directed along the normal to the device
plane~\cite{Kelly} at the interface.
Experimentally, in $GaAs$-$AsGaAl$ interface,  values for $ \alpha E_z
$ of  order $10^{-11}\; eV\; m$   were observed\cite{Nitta}, where
$\alpha=\frac{\hbar^2}{(2M_0c)^2}$.

\


Recently the "Rashba  interaction" led to an intense research
activity also including the effects of disorder and interaction.
The theory of transport in the presence of SO interaction
including disorder  was  developed
also in the presence of an in-plane magnetic field,
yielding a characteristic anisotropic
conductivity as a function of the magnetic field\cite{[1415]}.
More recently the so-called spin-Hall effect  was proposed
\cite{[10]} where the spin current response is due to an
applied transverse electric field.

The effects of a strong transverse magnetic field (i.e. applied in
the direction orthogonal to the 2DEG)  was also analyzed in
ballistic Quantum Wires in the presence of Rashba  coupling. In a
recent article we showed that the magnetic field enhances the spin
selection in the current and also gives very singular spin
textures in nanostructures \cite{[18]}.

In the last years also the case of a QD was analyzed: the effect
of spin-orbit coupling on  the electronic structure of
few-electron interacting QDs is  a suppression of Hund's rule due
to the competition of the Rashba effect and the exchange
interaction \cite{[28]}. This behaviour can be measured in
vertical QDs while, in lateral semiconductor QDs, weak
localization and universal conductance fluctuations  were analyzed
\cite{[30]}: in the presence of both SO scattering and a magnetic
field the conductance of a chaotic QD is a a function of the
parallel and perpendicular magnetic field and the SO coupling
strength.


In this paper we analyze what happens when a strong transverse
magnetic field acts on a vertical  QD in the presence of a Rashba
coupling. The small strength of the term $\alpha E_z$ allows for
a perturbative treatment of the SO coupling.
 We recall that, when a magnetic field is present,
 we should take into account also the Zeeman effect. However,
 because in $GaAs$ systems the band effect renormalizes the electron mass ($m*=0.067 M_0$)
 the Zeeman splitting is reduced  by a factor $4$, so that it  does not drive
any spin polarization in these systems. We will  apply our results
to two QDs of different radius, in order to emphasize the
competitive effects of the magnetic field, the SO perturbation and
 the electron electron interaction, and hence be able to define the
 hierarchy of the states belonging to  a
 multiplet.

\

In sec.II we discuss  the  the single particle problem by showing
the relevance of a term in the Rashba hamiltonian neglected in a
recent article about this same topic\cite{arturo,nota21}.

In sec.III we discuss the many electron case. In the first
subsection we report an overview of the phenomenology without the
SO coupling.
 In the second subsection  we investigate how the Rashba
coupling split  a fully spin polarized multiplet for a $5$
electrons QD.

In appendix A we discuss in detail the case of an unperturbed QD
with an arbitrary number of electrons, while in appendix B we
discuss the simplest case of a 2 electrons QD.

\section{Single electron}

Usually  a QD\cite{noicb}  is modelized by a two dimensional
harmonic confining potential $V({\bf r})=\frac{m}{2}\omega_d^2
|{\bf r}|^2$ which  {suggests}, with its symmetries, the choice of
the symmetric gauge  (${\bf A}=(-By/2,Bx/2,0)$).

The unperturbed hamiltonian for a single electron
$$
H_{T}=\frac{1}{2m}(\vec{p}-\frac{e}{c}\vec{A})^2+\frac{m
\omega_d^2}{2}r^2= \frac{\vec{p}^2}{2m}+\frac{m
\omega_T^2}{2}r^2-\frac{\omega_c}{2}L_z,
$$
 with $\omega_c=\frac{eB}{m c}$, commutes with $\hat{L}_z$ and
yields the usual Fock-Darwin\cite{FockDar} single particle energy
levels and the related eigenfunctions
\begin{eqnarray}\label{h12}
\varepsilon_{n_+,n_-}= \hbar n_+ \omega_- +
 \hbar n_- \omega_+ + \hbar \omega_T,
\end{eqnarray}
where $\omega_\pm=\omega_T \pm \frac{\omega_c}{2}$ and
$\omega_T=\sqrt{\frac{\omega_c^2}{4}+\omega_d^2}$.


In a  vertical  geometry, orbital  effects  induced by a
magnetic  field orthogonal to the dot  are dominant: the
increasing magnetic field gives an orbital polarization, up to a
state where all the electrons are in the lowest Pseudo Landau
Level (PLL) ($n_-=0$). In general the  PLLs  correspond to a fixed
 $n_-$ and become  the Landau Levels (LLs) in the
limit $\frac{\omega_d}{\omega_c}\rightarrow 0$. Because the
energy levels, when $\omega_d$ vanishes, depend just on $n_-$, the
LLs are infinitely degenerate while the PLLs are spaced by
$\omega_-$.


The SO hamiltonian in eq.(\ref{H_SO}) can be expressed in terms of
$\hat{a}_\pm$,$\hat{a}^\dag_\pm$ operators\cite{messiah}
\begin{eqnarray}\label{hmf2}
H_{SO}&=& i \frac{\vartheta_R}{\hbar}\sqrt{\frac{\hbar \omega_T}{2
m}}
\left(a_-^\dag(1+\gamma)-a_+(1-\gamma)\right)\sigma_+ \nonumber \\
&+&i \frac{\vartheta_R}{\hbar}\sqrt{\frac{\hbar \omega_T}{2 m}}
\left(a_+^\dag(1-\gamma)-a_-(1+\gamma)\right) \sigma_- ,
\end{eqnarray}
where $\vartheta_R={m \alpha E_z }$ and
$$
\gamma=\sqrt{\frac{\omega_c^2}{4 \omega_d^2+\omega_c^2}}.
$$
Thus, we can obtain the perturbation to the LLs by setting  $\gamma=1$ in
eq.(\ref{hmf2}) and the QD without magnetic field in the opposite
limit, i.e. $\gamma=0$.

The SO perturbation breaks the symmetries of the QD, so that $m$
and $s_z$ are not any
more good quantum numbers for the electron, while eq.(\ref{hmf2}) 
still preserves $j_z$.

For a single electron in the lowest PLL the spin-orbit correction
to the energy are, to the second order in $\vartheta_R$,
\begin{eqnarray}\label{pert0}
\Delta \varepsilon_{m,0,\downarrow}&\approx&
\alpha(\omega_c)^2\left\{\frac{(j_z+1/2)} {\hbar \omega_-}-
\left(\frac{1+\gamma}{1-\gamma}\right)^2 \frac{1}
{\hbar\omega_c}\right\}\\  \Delta
\varepsilon_{m,0,\uparrow}&\approx& -\alpha(\omega_c)^2
\frac{(j_z+1/2)} {\hbar\omega_-}, \label{pert01}
\end{eqnarray}
where
$$
\alpha(\omega_c)=\frac{\vartheta_R}{\hbar}\sqrt{\frac{\hbar
\omega_T}{2 M_0}}(1-\gamma).
$$
In the typical QDs that we consider, $\alpha(0)\approx
0.2-0.5 meV$.

From eq.(\ref{pert0}) and eq.(\ref{pert01}) we can conclude how
the two unperturbed spin-degenerate levels in the lowest
PLL are split by the SO perturbation. It is easy to show that the
states with spin down ($m,\downarrow$) have   energies lower than
those of the corresponding spin up states ($m,\downarrow$), at both
the low and the high limit  of the magnetic
field, although there might exist an intermediate regime where the
spin up states have the lowest energies.

It is also easy to calculate  the perturbed single particle
eigenfuctions, $\psi({\bf r})$, and then
 evaluate  the local spin density (LSD), as the mean value
 of the spin vector as a function of the position ${\bf r}$, i.e. ${\bf S}({\bf
r})=\langle\psi({\bf r})|{\bf S}|\psi({\bf r})\rangle$. {It could
also  be interesting to analyze the magnetization ${\bf
m}(r,\varphi)=\frac{{\bf S}(r,\varphi)}{\rho(r,\varphi)}$
, where $\rho$ is the charge density.
 The analysis of the
magnetization gives a better picture of the spin texture than the
one obtained from the LSD, because it describes the local
orientation of the spin vector by neglecting the attenuation of
the charge density.

The spin texture could be compared with the one which we showed
for a Quantum Wire\cite{[18]}. The spin has to be  orthogonal to
the velocity (current)  at each time as we found in
ref.\cite{[18]}, but it can reverse its direction going from the
center to the edge of the QD. We do not discuss in details this
behaviour, but just limit ourselves to point out the relevance of
the perturbative term neglected in ref.\cite{arturo}, where a
gauge noninvariant treatment was implemented.

\section{Many electron dot}

The many body hamiltonian  has in general the form
\begin{eqnarray}\label{h0d}
 \hat{H_0}=\sum_\alpha^\infty \varepsilon_\alpha \hat{n}_\alpha+
\frac{1}{2} \sum_{\alpha,\beta,\gamma,\delta}
V_{\alpha,\beta,\gamma,\delta}\,\,
\hat{c}^\dag_{\alpha}\hat{c}^\dag_{\beta}\hat{c}_{\delta}\hat{c}_{\gamma},
\end{eqnarray}
where  $\alpha\equiv(n_+,n_-,s)$ denotes the single particle state.
In the single particle energy level $\varepsilon_\alpha$,
 ${\hat{c}}^\dag_{\alpha}$ creates a particle in the state $\alpha$ and
$\hat{n}_\alpha\equiv \hat{c}^\dag_{\alpha}\hat{c}_{\alpha}$ is
the occupation number operator.

Because of the symmetries of eq.(\ref{h0d}) due to the properties
of electron electron interaction, once the number of
electrons in the QD ($N$) is fixed, we can characterize the ground state
(GS) with its spin $S$ and angular momentum $M$ and, starting from
eq.(\ref{h0d}), calculate in Hartree Fock approximation (HF) the
properties of the GS when the magnetic field $B$
increases\cite{klein,notahf}.

In the special case of just one filled PLL ($n_-=0$),  we can
introduce a semplified hamiltonian,
 which has many analogies with a  chiral Luttinger model, with
just one branch involved. In this case, we substitute
$V_{\alpha,\beta,\gamma,\delta}$ with  constant coupling
strengths, $g^\parallel$ (corresponding to a scattering process
involving electrons with the same spin) and $g^\perp$ (corresponding
to a scattering process involving electrons with opposite spins)
\begin{eqnarray}\label{h0}
  &H^\alpha_0&= \hbar v_F\sum_{m,s}\;m\;
c^{\dag}_{{m},s} c_{{m},s} \nonumber
\\
&+&
\sum_{m,\mu,q}\sum_{s,\sigma}\frac{g^\parallel}{\ell}\delta_{s,\sigma}\left(
c^{\dag}_{m+q,s} c^{\dag}_{\mu-q,\sigma}
c_{m,s}c_{\mu,\sigma}\right) \nonumber
\\
&+&
\sum_{m,\mu,q}\sum_{s,\sigma}\frac{g^\perp}{\ell}\delta_{s,-\sigma}\left(
c^{\dag}_{m+q,s} c^{\dag}_{\mu-q,\sigma}
c_{m,s}c_{\mu,\sigma}\right).
\end{eqnarray}
Here there appear both a field dependent interaction parameter
($1/\ell \equiv\sqrt{\omega_T/\omega_d}$) and   a field dependent
Fermi velocity $v_F=\omega_-$.

The hamiltonian in eq.(\ref{h0}) can be diagonalized
analytically and gives many interesting results. The effects of
the long range interaction, that sometimes can be quite important,
were discussed in a recent paper\cite{noicb} where we report some
details about the derivation of  eq.(\ref{h0}).

\subsection{Unperturbed Dot}

When there is no SO coupling, the
 quantum
numbers labeling the dot energy levels are the number of electrons
$N$, the total orbital  angular momentum along $z$, i.e.  $M$, the
total spin $S$ and the
 $z-$component of
the spin $S_z $.

The  orbital polarization, corresponding to all the electrons in
the lowest PLL, was often revealed in the past: if the magnetic
field is above a threshold value $B_s$ (depending on the number of
electrons $N$), a "singlet" ($S=0$ or paramagnetic state) is
observed where $N/2$ single electron states in the lowest PLL are
doubly filled.
As it was clear in the measurements for a $24$ electrons
dot\cite{klein}, by increasing the magnetic field
 $B$ above
a larger threshold value $B_p$, both $M$ and $S $ increase. The
electron system is expected to be  polarized, if the reduction in interaction
 energy (Coulomb exchange)  due to creating a finite spin polarization state exceeds
the cost in single particle kinetic energy.
 This spin polarization
phase ends when  $B= B^*$ and  a fully spin
 polarized (FSP),  $M=N(N-1)/2$ and $S=N/2$,
state is reached. By further increasing $B$, the electron density
reaches a maximum value, so that  the QD is in a state usually
called the {\it maximum density droplet}
(MMD)\cite{oosterkamp,mcdonald}. For larger $B$ values, the FSP
 state is disrupted: changes in the density are produced at the edge of the dot and  a
situation  known  as {\it dot reconstruction}\cite{klein2} occurs
by creating a  ring of filled states separated from a remaining
core of filled
 states by a ring of empty states.

 \begin{figure}
\includegraphics*[width=1.0\linewidth]{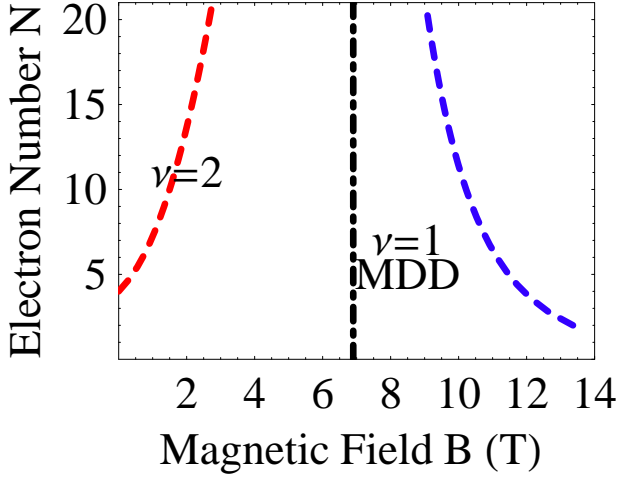}
 \caption{{
We start  from the
 singlet ($\nu=2$) threshold field (left red line) where the droplet is in  a non polarized
 (paramagnetic) state;
when the field $B$ grows, the  electron system is expected to be
spin polarized, if the reduction in interaction
 energy (Coulomb exchange)  due to creating a finite spin polarization state exceeds
the cost in single particle kinetic energy, until the total
polarization  state ($\nu=1$ FPS at  $B>B^*$, black line),
sometimes called the maximum density droplet (MMD). For this state,
all single particle states with $m$ less than $N$ have just one
electron, while all others are empty. In this regime, HF predicts that a
reconstruction of the MDD occurs by creating a ring of filled
states, separated from a remaining core of filled
 states by a ring of empty states, for fields above the blue line.
Here $U \approx 0.1-0.15 \hbar \omega_d$ and the results can be
compared with the ones in ref.\cite{oosterkamp}. Further explanations about the calculations
of the phases in the $B-N$ plane are reported in appendix A. }}
\end{figure}

The theoretically obtained diagram in Fig. (1),  usually
reported\cite{klein,oosterkamp} as phase diagram, where  the
 spin properties of the GS of the many electron system are shown
 as a function of
the number of electrons ($N$) and the magnetic field ($B$), is
calculated in appendix A  and can also be compared with
measurements in \cite{oosterkamp}. In this calculation, the long
range effects of the interaction are sometimes quite
relevant\cite{noicb}.

\subsection{Introduction of the SO}

Here we follow the perturbative approach  in the SO coupling, in
order to calculate the energy splitting of the FSP multiplet.

 In
the presence of SO coupling, $J_z = M + S_z $ becomes the good quantum
 number and  the FSP, which is the GS of the unperturbed dot  if $B$ is above a threshold value,
 corresponds to a multiplet $J_z= N(N-1)/2+S_z$.

In order to explain what happens when the SO coupling acts on the FSP, we have
to put the SO hamiltonian eq.(\ref{hmf2}) in the second
quantization form
\begin{eqnarray}\label{hmf3}
H_{SO}&=& i \alpha(\omega_c)\sum_{m}  \left(c^\dag_{1,m,\uparrow}c_{0,m,\downarrow}-c^\dag_{0,m,\downarrow}c_{1,m,\uparrow}\right)\frac{(1+\gamma)}{(1-\gamma)} \nonumber \\
&-&\left(c^\dag_{0,m-1,\uparrow}c_{0,m,\downarrow}-c^\dag_{0,m+1,\downarrow}c_{0,m,\uparrow}\right)\sqrt{j_z+1/2}.
\nonumber
\end{eqnarray}
Here $c^\dag_{n_-,n_+,\sigma}$ are the usual creation
operators for the electrons,  $m$ labels $n_+$, so that $m$ is the
angular momentum if $n_-=0$ (i.e. if the state belongs to the
lowest PLL).

 Now we want to calculate the energy splitting of the states
belonging to a multiplet, starting from the wavefunction
$\Psi^0_{N,M,S,S_z}$, where $-S\leq S_z \leq S$, up to  the second
perturbative order.
 Then, we take in account all the Slater determinants with $N$ electrons
$\Psi^{SD}_{N,J_z,l}$, where $l$ labels each different Slater
determinant. We calculate the perturbative energy for the states
$\Psi^0$
$$
\Delta
\varepsilon_{N,M,S,S_z}=\sum_{l}\frac{{|\langle\Psi^{SD}_{N,J_z,l}|H_{SO}|\Psi^0_{N,M,S,S_z}\rangle|}^2}{
\varepsilon_{N,M,S,S_z}- \varepsilon_{l}},
$$
where
$$\varepsilon_{l}=\langle\Psi^{SD}_{N,J_z,l}|H_{0}|\Psi^{SD}_{N,J_z,l}\rangle.$$
Starting from the hamiltonian in eq.(\ref{h0}), the splitting
energies for the FSP multiplet can be analytically expressed in
terms of $S,S_z$ and  $U=(g^\perp-g^\parallel)$
\begin{eqnarray}\label{pert}
 \Delta \varepsilon_{N,S_z}(U,\omega_c)&\approx& -\alpha(\omega_c)^2\{
  \frac{
(S+S_z)(S+S_z+1)} { \sqrt{\frac{\omega_T}{\omega_d}}U\left[S^2 - (S_z-1)^2\right]+\omega_-}\nonumber \\
&+& \left(\frac{1+\gamma}{1-\gamma}\right)^2 \frac{ (S - S_z)}
{\sqrt{\frac{\omega_T}{\omega_d}}U\left[S^2 - (S_z+1)^2\right]+\omega_c}\nonumber \\
&+&  \frac{ (S-S_z)(S-S_z-1)} {
\sqrt{\frac{\omega_T}{\omega_d}}U\left[S^2 - (S_z+1)^2\right]-\omega_-}\}.
\end{eqnarray}

 The SO interaction lifts the spin degeneracy.
From a theoretical point of view, we can analyze  how the hierarchy
of the states belonging to  a
 multiplet, from ground to top, is due to the competitive effects of the magnetic field, the SO perturbation  and
 the electron electron interaction.
So, we can discuss two regimes, i.e. the one of weak SO (WSO)
perturbation and the opposite regime, i.e. the weak interaction
(WI) one In fact,  the strength
of $U$ is responsible 
for the ordering in energy of the sequence $J_z$, in competition
with the magnetic field.

 The two states which compete as GS in
the $(2S+1)$-plet are the one with $S_z=-S$ in the  WI regime and
the one with $S_z=S-1$ in the WSO limit. The magnetic field favors
$J_z=J_{min}$ as the GS if $U$ is not too large. This could also
be shown by treating  the electron electron interaction as a
perturbation. Our results confirm the hierarchy of the multiplet
states from the GS $M-S$ to the rather unperturbed $M+S$. These
results are shown in Fig.(2).
\begin{figure}
 \includegraphics*[width=.45\linewidth]{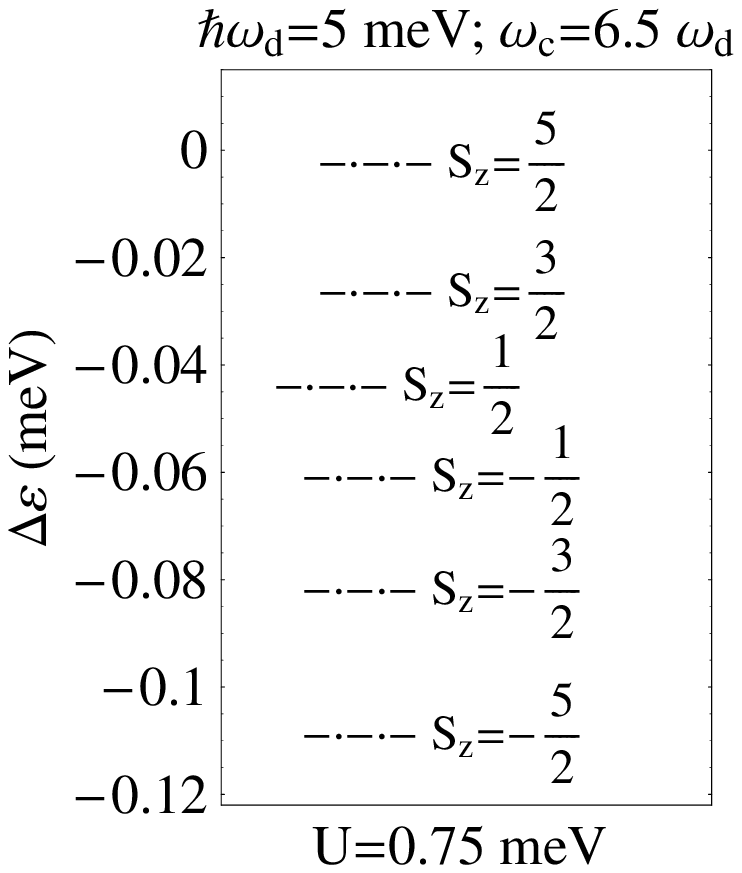}
\includegraphics*[width=.45\linewidth]{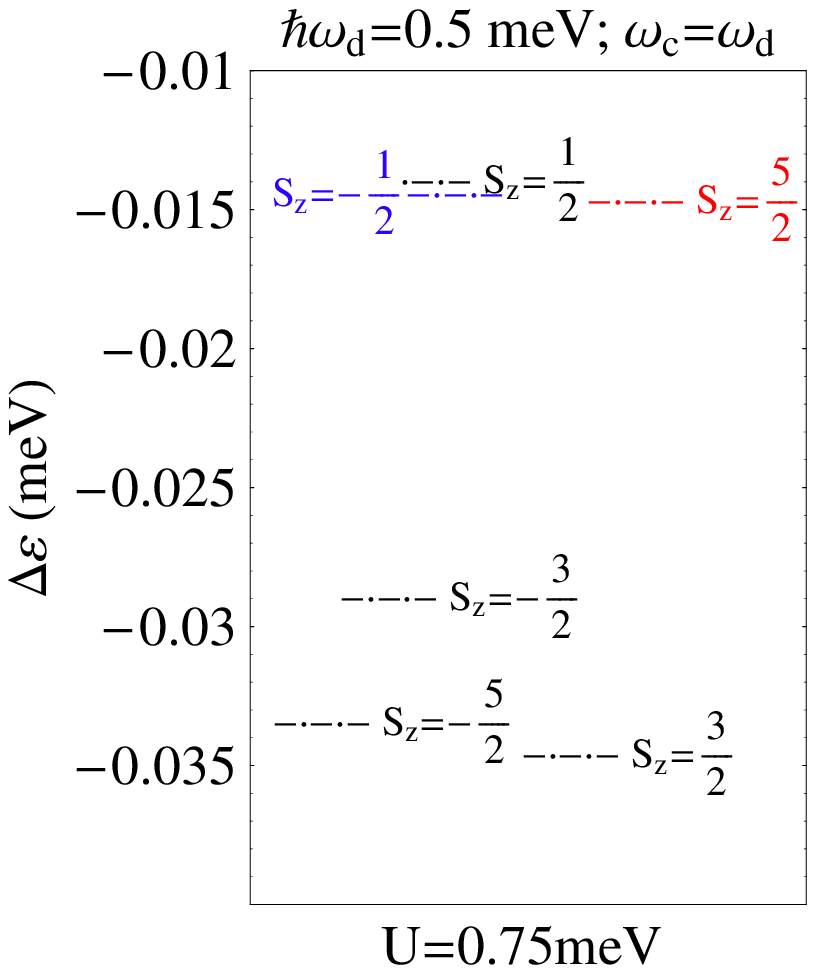}
\caption {The energy splitting for two $5$-electrons QDs of different radius in the FPS.
 On the right panel the radius is $\sim 3$ times larger than on the left, yielding a reduction
 of $\omega_d$ by a factor $\sim 9$.  } \end{figure}

In the opposite limit, the WSO shown in Fig.(2,right),   the
presence of the state with $J_z=J_{max}-1$ as a GS could seem
suspicious; however it is a consequence of the strong reduction in
the interaction due to the spin polarization. In fact, all the
states obtained from it by applying the perturbation are fully
polarized states ($S=N/2$) and strongly reduce the interaction
energy. This is better explained in appendix B,
where the energy levels of a two electrons dot are discussed.

{ Both  cases discussed above  yield  a singular spin texture, due
to the combined effects
 of the interaction responsible for the polarization and the Rashba
 effect.}

\begin{figure}
\includegraphics*[width=.495\linewidth]{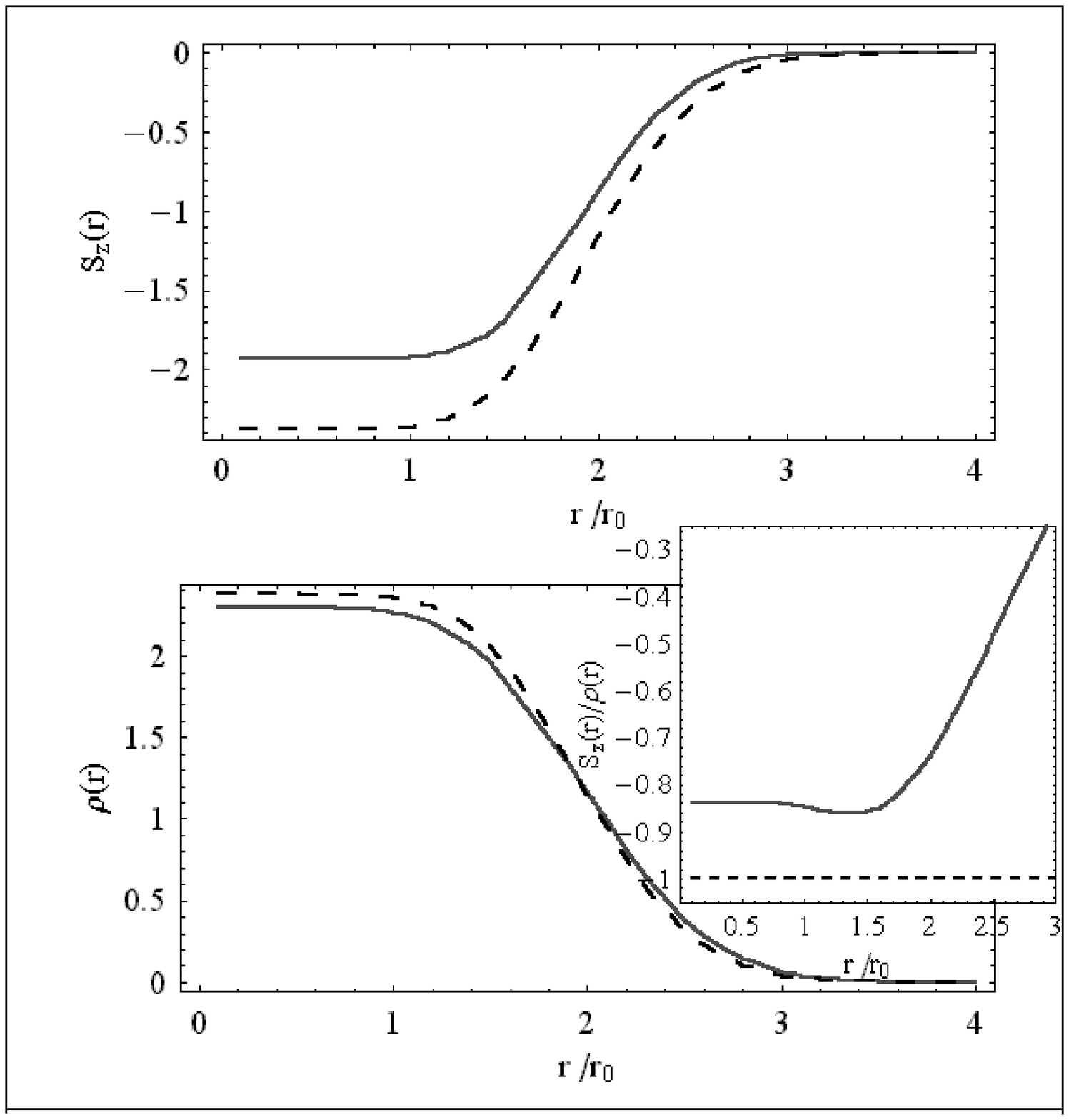}
\includegraphics*[width=.475\linewidth]{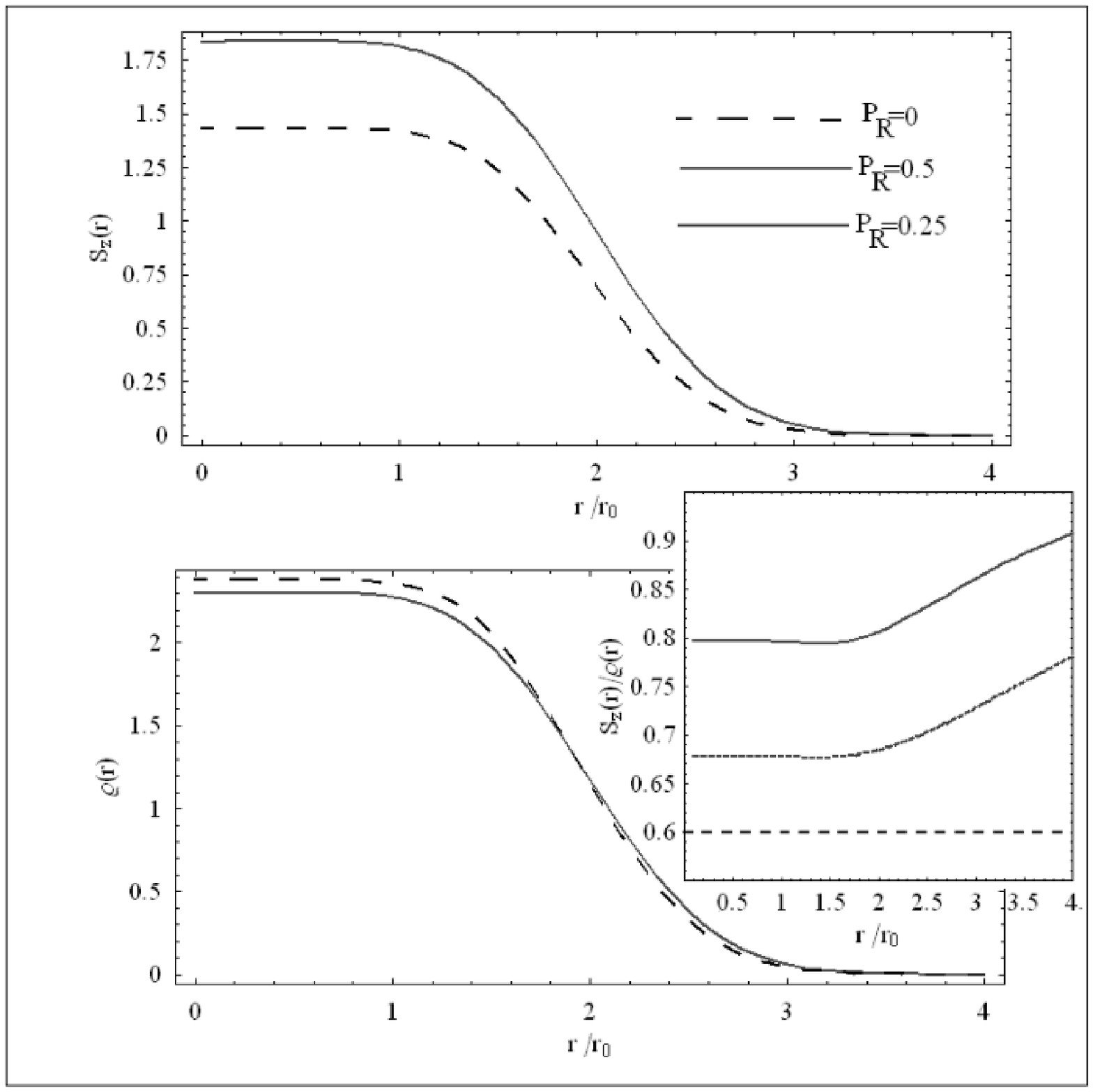}
 \caption { Azimuthal Local Spin
Density $S_z$ (top), Charge density $\rho$ (bottom) and
magnetization $m_z=S_z/\rho$ (insets) as a function of the radius,
for different strengths of the SO coupling $\alpha(0) \propto p_R$:
(left) the state $S_z=-5/2$, (right)  the state ($S_z= 3/2$). Here
$r_0=\sqrt{\hbar/(m\omega_T)}$.  }
\end{figure}


\section{conclusions}

In this article we discussed how the presence of a SO coupling
affects the charge and the spin polarization in a vertical
disk-shaped quantum dot  under a strong magnetic field.

 When there is no Rashba
coupling,  the  combined effect  of the  electrostatic forces
between the $N$ electrons confined in the QD and the Pauli
principle induces a spin polarization, in the presence of a strong
magnetic field $B$. However, also in this case  the electron
system preserves the quantum number  $S$.

The presence of
 an electric field parallel to $B$, coupled to  the spin $S$ by a Rashba
 term, breaks this symmetry.
Here we have analyzed
 the splitting in the
 levels of the states belonging to a
 multiplet, and their hierarchy from ground to top.

Now we can discuss the effect of this coupling for two different
real QDs filled by $5$ electrons, by choosing the values of the
parameters so that the FSP multiplet is  the GS of the unperturbed
QD.

The first one has $\hbar \omega_d\approx 5 meV$, so that the
threshold field for having the FPS, $B^*$, corresponds to
$\omega_c \lesssim 6.5 \omega_d$ ($B\approx 6.5 T$). The strength
of the interaction ($g^\perp\approx 3 meV$) yields $U\approx 0.5/1
meV$. In this case we predict a phase diagram quite similar to the
one measured in ref.\cite{oosterkamp} (see Fig.(1)), while the
hierarchy of the split FPS multiplet at $\omega_c =6.5  \omega_d$
favors $S=-5/2$ as GS.

The second QD has a radius 3 times larger than the first one, so
that $\hbar \omega_d\approx 0.5 meV$. In this case the threshold
$B$ corresponds to $\omega_c \lesssim \omega_d$, while the
parameter $U$ has to be unchanged. For this secon QD we predict a
different hierarchy of the split FPS multiplet at $\omega_c
=\omega_d$, which favors $S=3/2$ as GS. The comparison between the
two cases is shown in Fig.(2).

 Our results  can be compared with the ones obtained in a recent paper\cite{arturo},
  where another term of the SO
perturbation dominates, obtained by breaking gauge invariance.
Hence, we believe  that the hierarchy of the states reported  in
ref.\cite{arturo} has to be revised, by including the correct gauge
invariance term. We also believe that the inclusion of all the
terms in the perturbation modifies the charge density and the LSD,
as can be seen in Fig. (3).
It could have some hard consequences on the analogy with the
"Skyrmion states" discussed in ref.\cite{arturo}.

\acknowledgements

 \noindent This work is partly supported by the Italian
Research Ministry MIUR, National Interest Program under grant
COFIN 2002022534.

\
\appendix
\section{Spin phases in QDs}
\label{a1}
Here we want to illustrate some simple calculations that allow us
to deduce at which value of the magnetic field some phenomena
occur. This kind of simplified calculations allows for a  good
knowledge of the phase diagram shown in Fig.(1). This approach
assumes the electron electron interaction as a perturbation with
respect to the kinetic energy and is stopped at first order, so
that no correlation effects are appreciable in this way.

We can fix the number of electrons $N$, write the single Slater
Determinants (SDs) $\Psi^{SD}_{N,M,S,S_z}$ and then compare the
mean value of the hamiltonian eq.(\ref{h0}) on these
wavefunctions. We can write, in the Lowest PLL limit,
 \bea
|M,S_z \rangle = \prod _{m ,\mu=0}^{\infty} \left
(\hat{c}^\dagger_{0,m,\uparrow}\right )^{n_{0,m,\uparrow}}
\left(\hat{c}^\dagger_{0,\mu,\downarrow}\right)^{n_{0,\mu,\downarrow}}
|0 \rangle , \label{sld} \eea where
$$
M=\sum _{m=0}(n_{0,m,\uparrow}+n_{0,m,\downarrow})m \;\;\;\;
S_z=\frac{1}{2}\sum _{m=0}(n_{0,m,\uparrow}-n_{0,m,\downarrow}).
$$
In the Lowest PLL limit some of these SDs, i.e. those with  fixed $M$,
$S$ and $S_z=\pm S$, are eigenstates of the hamiltonian, because
no other states with the same $M$ and $S_z$ can be found. As the eigenstates are not
degenerate, there are no correlation effects, hence the HF approach yields
exact results. This is
the case of the singlet, $\nu=2$ state, the FSP filling $\nu=1$, but
also the state of their fundamental excitations.

For singlet, spin-degenerate states belonging to the lowest
PLL\cite{oosterkamp}} (Singlet, $\nu=2$) one has

$M=N/2(N/2-1)$ $S=S_z=0$.

As $B$ is increased further, it becomes energetically
favorable for an electron to flip its spin and move to the edge of
the dot\cite{oosterkamp} (First spin flip)

$M=N/2(N/2-1)+1$ $S=S_z=1$.

Then the electrons flip the spins one by one from the edge
to the core\cite{klein}, until the last spin flip from the state

 $M=N(N-1)/2-N+1$
$S=S_z=N/2-1$

to the Fully Polarized state (FSP or MDD, $\nu=1$)

$M=N(N-1)/2$ $S_z=S=N/2$.

When B is increased further, the angular momentum states
shrink in size such that the density of the MDD increases.  At
some threshold B value  the direct Coulomb interaction has become
so large that the MDD breaks apart into a larger, lower density
droplet (LDD)."\cite{oosterkamp} For larger $B$ values, the FSP
 state is disrupted: changes in the density are produced at the edge of the dot and  a
situation  known  as {\it dot reconstruction}\cite{klein2} occurs
by creating a  ring of filled states separated from a remaining
core of filled
 states by a ring of empty states

(Edge hole-particle) $M=N(N-1)/2+1$ $S_z=N/2$ .

The Luttinger-like hamiltonian is very easy to solve, within this
approximation. For example,
 the mean value of the
hamiltonian  on the FSP is a function of the total spin \bea
\langle
M,S_z|H|M,S_z\rangle&=&\omega_-M
+\frac{N^2}{4}\left(g^\parallel+g^\perp \right) \nonumber
\\&+&\left(2g^\parallel-g^\perp \right)S_z^2-N g^\parallel.\label{fs} \eea
Thus, we can observe the symmetry $\pm S_z$, from which we deduce
that two degenerate state FSPs are possible with $S_z\pm N/2$  .

Also for the generic interaction we can calculate \bea
\langle
M,S_z|H|M,S_z\rangle&=&\omega_-M\nonumber \\
+ \sum_{m
,s}\sum_{\mu,\sigma}\left(g^\parallel_{m,\mu}\delta_{s,\sigma}+g^\perp_{m,\mu}\delta_{s,-\sigma}
\right)n_{0,m,s} n_{0,\mu,\sigma}.\label{fs2} \eea

\subsection{Charge and spin phases in an unperturbed QD}

Now we can calculate the energy of each SD and, by  comparison, we
can evaluate three threshold (critical) fields corresponding to
the different phases. So, we can write analytical formulas for the
threshold fields depending on the number of electrons in the dot
(some of these formulas are obtained in the limit of large
magnetic field $\omega_-\approx \omega_d^2/\omega_c$ and
$\omega_T\approx \omega_c/2$). The orbital polarization, i.e. the
condition for having all the electrons in the lowest PLL, reads
\bea\label{a1}
 \omega_-\left(\frac{N}{2}-1\right)<\omega_+\Rightarrow
\omega_c>\omega_d\sqrt{\frac{(N-4)^2}{2(N-2)}}.
\eea

The second threshold corresponds to  the interaction-induced spin
flip at the edge \bea\label{a2}
\omega_-<\sqrt{\frac{\omega_T}{\omega_d}}U \Rightarrow
\omega_c>\omega_d\left({\frac{2\omega_d^2}{
U^2}}\right)^\frac{1}{3}. \eea In this model all the spins flip at
the same critical field, provided the magnetic field is above the
total orbital polarization field obtained from the condition
\bea\label{a3} \omega_-\left(N-1\right)<\omega_+\Rightarrow
\omega_c>\omega_d\sqrt{\frac{(N-2)^2}{(N-1)}}. \eea

Now we are ready to plot the phase diagram as shown in Fig.(1). {
The equations above allow us to predict the magnetic fields
corresponding to the transition $\nu=2 \rightarrow \nu <2$. In
agreement with the experimental data of ref.\cite{oosterkamp}we
predict $\omega_c=0$ for $N=4$. By eq.(\ref{a1}) we can also
deduce the value of the magnetic field for having the
transition  $\nu=2 \rightarrow \nu <2$ for fixed  $N$ and $B$. The
experiment in ref.\cite{oosterkamp} also show the transition to
FSP ($\nu>1 \rightarrow \nu =1$) at a field $B\approx 6 T$. This
transition can be predicted in our model (eq.(\ref{a2})) by
putting $U=(g^\perp-g^\parallel)\approx 0.1-0.15 \hbar \omega_d$.

Although the  model gives good results in the description of these
two transitions in the limit of not so large $N$,  from the
experimental data in ref.\cite{oosterkamp} we can also deduce that
at very strong field the FSP can be disrupted by the formation of
hole particle excitations at the edge. In order to reproduce this
phenomenon, we have to introduce
 a modification of the interaction
parameters, $g^\parallel_{m,\mu}$ and $g^\perp_{m,\mu}$, by taking
in account their dependence on the quantum numbers of the
interacting electrons  $m$, $\mu$. So we can assume, starting from
a Dirac $\delta$ model of the interaction,
$$
g^i_{m, m',\mu ,\mu '}=g^i \delta_{m+\mu,m'+\mu'}
\frac{2^{-{m+\mu}} \Upsilon_{m+\mu}}{\Upsilon_m\Upsilon_\mu
\Upsilon_{m'} \Upsilon_{\mu'}},$$ where $
\Upsilon_m=\sqrt{\frac{\Gamma[1+m]}{2}}
$ and $i=\perp,\parallel$.

A second effect, that the constant interaction model is not able
to explain, concerns the transitions from $\nu=2$ to $\nu=1$
observed in the experiments in ref.\cite{klein} for a $24$
electrons Dot. The analysis about the phase corresponding to
$2>\nu>1$ is one of the central points of the discussion in
ref.\cite{klein}, where experimental data were compared with
predictions obtained by using the HF approach.

In this case our model with constant interaction can just
predict, starting from the values of the two transitions fields
$B_{\nu=2}\approx 1.5 T$ and $B_{\nu=1}\approx 4.5 T$, that the QD
has a radius larger than the one used in the previous experiment
($R/r\approx 2-3$) and a different interaction strength.  The
modified interaction parameters allow us to calculate also the
spin susceptibility, obtaining results in agreement with the
ones reported in ref.\cite{klein}. }

\section{2 electrons dot}

In order to explain why $J_{max}-1$ can be the GS for a strongly
interacting electron system, we can discuss what happens for the
simplest case of a $2$ electrons QD. The unperturbed state  is
the triplet ($M=1$ $S=1$)
$$
|a\rangle=c^\dag_{0,0,\uparrow}c^\dag_{0,1,\uparrow}|0\rangle $$
$$
|b\rangle=\frac{1}{\sqrt{2}}\left(c^\dag_{0,0,\uparrow}c^\dag_{0,1,\downarrow}+c^\dag_{0,0,\downarrow}c^\dag_{0,1,\uparrow}\right)|0\rangle
$$ $$
|c\rangle=c^\dag_{0,0,\downarrow}c^\dag_{0,1,\downarrow}|0\rangle .
$$ The SO hamiltonian  in the second quantization form, if we
consider the limit of strong magnetic field (a condition enforced
because the $FSP$ states need a magnetic field above a threshold), reads
\begin{eqnarray}\label{hmf3}
H_{SO}&\approx& i \alpha(\omega_c)\sum_{m}
\left(c^\dag_{1,m,\uparrow}c_{0,m,\downarrow}\right)\frac{(1+\gamma)}{(1-\gamma)}.
\nonumber
\end{eqnarray}
It follows that the $H_{SO}|a\rangle=0$, so that the state $a$ is
unperturbed.

The state $|b\rangle$ has an interesting property, when the SO hamiltonian
acts on it: $|\beta\rangle=H_{SO}|b\rangle$ has total spin $S=1$.
In fact
$$H_{SO}|b\rangle= |b\rangle+i \alpha(\omega_c)
\left(c^\dag_{1,0,\uparrow}c^\dag_{0,1,\uparrow}
+c^\dag_{0,0,\uparrow}c^\dag_{1,1,\uparrow}\right)\frac{(1+\gamma)}{(1-\gamma)}|0\rangle$$
which, in a different formalism, could be written as
$$
H_{SO}|b\rangle=\rho_{M=1}({\bf r_1,r_2})\chi_{1,0}+i
\alpha(\omega_c)\rho_{M=0}({\bf r_1,r_2})\chi_{1,1}.
$$
This is a general property of the states with $J_{max}-1$, i.e.
under the action of $H_{SO}$ they preserve the total spin. So, we
can apply Hund's rule, in order to conclude that these states
minimize the interaction. This result does not depend either
on the formalism nor on the model of interaction.

We just finish by showing that the state $|c\rangle$ gives \bea
H_{SO}|c\rangle &=& \rho_{M=1}({\bf r_1,r_2})\chi_{1,-1}\nonumber
\\ &+& i \alpha(\omega_c)\rho_{M=0}({\bf
r_1,r_2})(\frac{\chi_{1,0}\pm\chi_{0,0}}{\sqrt{2}}).\nonumber \eea

So, we can conclude that, in a general case, the interaction splits
the perturbed  states of a multiplet by favoring the state
$J_{max}-1$, and this state can be the lowest energy state if
the interaction is quite strong, as we show in Fig.(2.right).

Because in our previous discussion we did not specify which kind of
interaction we used, these results are rather general. The central
question is: can $U$ be strong? If we take a model with constant
interaction (in configuration space) there is no Coulomb
exchange and $U$ is always 0. In general $U$ does not vanish, and
the factor $\sqrt{\frac{\omega_T}{\omega_d}}$ enforces its action
at strong values of the magnetic field.


\bibliographystyle{prsty} 
\bibliography{}

\begin{thebibliography}{99}
\bibitem{kou}
 L.~P.\ Kouwenhoven, C.~M.\ Marcus, P.~L.\ McEuen, S.\
  Tarucha, R.~M.\ Westervelt, and N.~S.\ Wingreen, in {\em Mesoscopic Electron
  Transport}, edited by L.L.Sohn, L.P. Kouwenhoven and G.Schon (Kluwer Axcademic Publisher, Series E, 1997).

\bibitem{klein}
O. Klein, C. de C. Chamon, D. Goldhaber-Gordon, M.A. Kastner and
X.-G. Wen, {\it Phase Transitions in Artificial Atoms, Quantum
Transport in Semiconductor Submicron Structures} NATO ASI Series
E, B. Kramer ed., p. 239, 1996; O. Klein, D. Goldhaber-Gordon, C.
de C. Chamon and M.A.~Kastner,
    \prb  {\bf 53},  4221  (1996).

\bibitem{oosterkamp}
T.H. Oosterkamp, J.W. Jansen, L.P. Kouwenhoven, D.G. Austing, T.
Honda, S. Tarucha,
 Phys. Rev. Lett. {\bf 82}, 2931 (1999).


\bibitem{spintro}
 D.D. Awschalom, D. Loss and N. Samarth, {\it Semiconductor Spintronics and
Quantum Computation } (Springer, Berlin, 2002);
      B. E. Kane, Nature {\bf 393}, 133 (1998).
      \bibitem{wolf}
 { {S.~A.} {Wolf}}
 D. D. Awschalom,  R. A. Buhrman,  J. M. Daughton,  S. von Molnar,  M. L. Roukes,
  A. Y. Chtchelkanova and  D. M. Treger, {Science} \textbf{{294}},
   {1488} ( {2001}).
\bibitem{Datta}
S. Datta and B. Das, Appl. Phys. Lett. {\bf 56}, 665 (1990).
\bibitem{nota}
For a review about the triangular well at the AsGa-As interface
see T.J. Thornton, Rep. Prog. Phys. {\bf 58}, 311 (1995).
\bibitem{[7]} Y.A. Bychkov and E. Rashba, JETP Lett. {\bf 39}, 78 (1984); J. Phys. C {\bf 17}, 6039 (1984).
\bibitem{[8]} G. Dresselhaus, Phys. Rev. {\bf 100}, 580 (1955).
\bibitem{Stormer} H.~L.~Stormer, Z.~Schlesinger, A.~Chang, D.~C.~Tsui,
A.~C.~Gossard, W.~Wiegmann, \prl {\bf 51}, 126 (1983).
\bibitem{Nitta} J.~Nitta, T.~Akazaki, H.~Takayanagi, \prl {\bf 78}, 1335
(1997).
\bibitem{Thankappan}  L.~D.~Landau, E.~M.~Lifshitz, {\it Quantum Mechanics}
(Pergamon Press, Oxford, 1991).

\bibitem{morozb}
 { {A.~V.}  {Moroz}}  {and}
   { {C.~H.~W.}  {Barnes}},
   {Phys. Rev. B} \textbf{ {61}},
   {R2464} ( {2000}).



\bibitem{Kelly} M.~J.~Kelly, {\it Low-dimensional semiconductors: material,
physics, technology, devices} (Oxford University Press, Oxford,
1995).
\bibitem{[1415]}R.  Raimondi, M. Leadbeater, P. Schwab, E. Caroti and C. Castellani, Phys.
Rev. B   {\bf 64}, 235110 (2001);  P. Schwab and R. Raimondi,  EPJB  {\bf 25}, 483 (2002).
\bibitem{[10]} J. Sinova, D. Culcer, Q. Niu, N. A. Sinitsyn, T. Jungwirth, A. H. MacDonald, cond-mat/0307663;
R. Raimondi and P. Schwab, cond-mat/0408233.
\bibitem{[18]}S. Bellucci and P. Onorato, Phys. Rev. B {\bf 68}, 245322 (2003).
\bibitem{[28]}M. Governale, Phys. Rev. Lett. {\bf 89}, 206802 (2002).
\bibitem{[30]}I.L. Aleiner, V.I. Fal'ko, Phys. Rev. Lett   {\bf 87}, 256801 (2001);  J.H. Cremers, P.W. Brouwer, V.I. Fal'ko,
Phys. Rev. B {\bf 68}, 125329 (2003).
\bibitem{arturo}P. Lucignano, B. Jouault, A. Tagliacozzo,   Phys. Rev. B {\bf 69},
045314 (2004).
\bibitem{nota21}
Note added. An improved treatment, which has completed the calculations in ref.\cite{arturo}, including the neglected terms of
the Rashba Hamiltonian,
has now been published in P. Lucignano, B. Jouault, A. Tagliacozzo and B. L. Altshuler  Phys. Rev. B {\bf 71}, 121310(R),
where a numerical diagonalization is performed for the system with few interacting electrons. This paper, presented
six months after our submission, also presents a discussion about the far-infrared radiation absorption of a
quantum dot with few electrons.
\bibitem{noicb}
S. Bellucci and P. Onorato,  Phys. Rev. B {\bf 71}, 075418 (2005).
\bibitem{FockDar}
V. Fock, Z. Phys. {\bf 47}, 446 (1928); C.G. Darwin, Proc.
Cambridge Philos. Soc. {\bf 27}, 86 (1930).



\bibitem{messiah}A. Messiah, {\it Quantum Mechanics} (North Holland Publishing Co., 1961).


 \bibitem{mcdonald}
A.H. Mac Donald, S.-R. E. Yang and M.D. Johnson, Aust. J. Phys. {\bf 46},  345  (1993).
\bibitem{klein2}
O. Klein, C. de C. Chamon, D. Tang, D.M.~Abusch-Magder, X.-G.~Wen, M.A.~Kastner and S.J. Wind,
    \prl  {\bf 74},  785  (1995).
\bibitem{notahf}
In ref.\cite{klein} the agreement between HF approach and
experiments is discussed with some details, and here we show (see
Fig.(1)) how our prediction about the phase diagram obtained in
ref.\cite{oosterkamp} could be compared with the experiment.
Obviously, our approach allows just a limited number of predictions,
when the level of degeneracy is low. We have to remark that our
approximation could be valid just in the limit of $U/\ell\ll
\hbar\omega_+$.
\end{thebibliography}

\end{document}